\documentstyle[aps,pre,psfig]{revtex}

\begin{document}
\draft

\twocolumn[\hsize\textwidth\columnwidth\hsize\csname @twocolumnfalse\endcsname

\title{Director Field Configurations around a Spherical Particle in a
Nematic Liquid Crystal}

\author{Holger Stark}
\address{Institut f\"ur Theoretische und Angewandte Physik, Universit\"at
Stuttgart, Pfaffenwaldring 57, D-70550 Stuttgart, Germany}

\maketitle

\begin{abstract}
We study the director field around a spherical particle immersed in a 
uniformly aligned nematic liquid crystal and assume
that the molecules prefer a homeotropic orientation at the surface of
the particle. Three
structures are possible: a dipole, a Saturn-ring, and a surface-ring
configuration, which we investigate by numerically minimizing the
Frank free energy supplemented by a magnetic-field and a surface term.
In the dipole configuration, which is the absolutely stable structure
for micron-size particles and sufficiently strong surface anchoring,
a twist transition is found and analyzed. We show that a
transition from the dipole to the Saturn ring configuration is induced
by either decreasing the particle size or by applying a magnetic field.
The effect of metastability and the occurence of hysteresis in connection
with a magnetic field are discussed. The surface-ring configuration
appears when the surface-anchoring strength $W$ is reduced. It is also
favored by a large saddle-splay constant $K_{24}$. A comparison with
recent experiments \cite{itapdb:Poulin1997,itapdb:Poulin1998}
gives a lower bound for W, {\em i.e.}, $W > 0.6 \, \text{erg}/\text{cm}^{2}$
for the interface of water and pentylcyanobiphenyl (5CB)
in the presence of the surfactant sodium dodecyl sulfate.
\end{abstract}


\vskip2pc]

\narrowtext

\section{Introduction} \label{sec intro}

Dispersions of particles in a host fluid are part of our everyday life
and an important matter of state. Since they appear in food 
({\em e.g.}\ milk), drugs, paints, and ink, they are of considerable 
technological importance.
In colloidal suspensions the particles are solid, whereas emulsions
are dispersions of liquid droplets coated with a surfactant. There
exists a whole zoo of interactions between the particles whose
delicate balance determines the stability of a dispersion. Besides
the conventional van-der-Waals, screened Coulombic, and steric interactions 
\cite{itapdb:Russel1995},
fluctuation-induced Casimir forces \cite{itapdb:Krech1994,itapdb:Mostepanenko1997} and depletion forces \cite{itapdb:Russel1995,itapdb:Dinsmore1996,itapdb:Dinsmore1998,itapdb:Rudhardt1998} have
attracted a lot of interest.

When the particles are dispersed in a nematic liquid crystal, {\em i.e.},
an anisotropic fluid, where
elongated organic molecules are aligned on average along a common
direction called director, additional long-range forces due to
elastic deformations of the director field are induced. 
These forces are of dipolar or quadrupolar type depending on the 
symmetry of the director configuration around the particle 
\cite{itapdb:Brochard1970,itapdb:Ramaswamy1996,itapdb:Ruhwandl1997,itapdb:Poulin1997,itapdb:Lubensky1998}, and they were confirmed
by recent experiments in inverted nematic emulsions
\cite{itapdb:Poulin1997,itapdb:Poulin1997a,itapdb:Poulin1998}. 
On the other hand topological point defects in the orientational order
of the molecules give rise to a short-range repulsion 
\cite{itapdb:Poulin1997,itapdb:Poulin1998}.
Related early work deals with forces between walls
mediated by an enclosed nematic liquid crystal \cite{itapdb:Horn1981,itapdb:Poniewierski1987a}. Surface-induced nematic order in the
isotropic phase close to the isotropic-nematic phase transition leads to
a short-range interaction as reported in most recent experimental and 
theoretical work \cite{itapdb:Musevic1996,itapdb:Borstnik1997,itapdb:Borstnik1998,itapdb:Galatola1998}. Even Casimir forces arising
from fluctuations in the liquid crystalline order parameter have been
calculated \cite{itapdb:Ajdari1991,itapdb:Swanson1995,itapdb:Ziherl1998}.

Before treating the interaction between particles it is necessary
to understand how one single particle behaves in a nematic environment.
This article deals with the possible director configurations around a 
spherical particle, which is placed into a uniformly aligned nematic 
liquid crystal, and which prefers a homeotropic, {\em i.e.}, perpendicular
anchoring  of the molecules at its surface. This is already a
complicated problem. Since it cannot be solved analytically
without employing an ansatz function for the director configuration
and further approximations, we will handle it by numerical methods.

\begin{figure}
\centerline{\psfig{figure=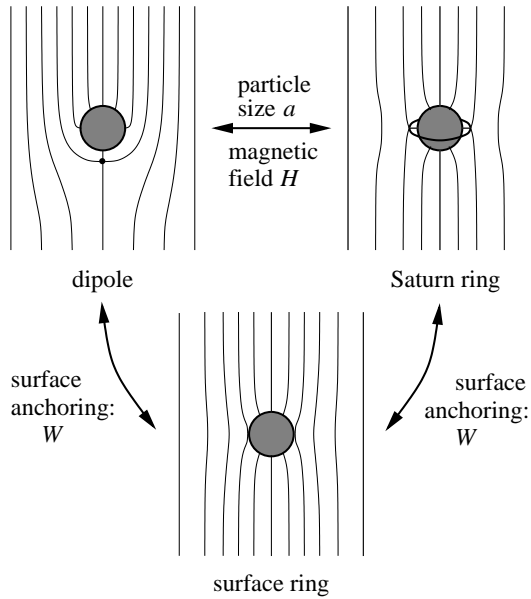,width=7cm}}

\vspace{0.5cm}

\caption[]{A spherical particle with a preferred homeotropic anchoring
at its surface that is placed into a uniformly aligned nematic liquid
  crystal exhibits three possible structures: the dipole configuration
  where the particle is accompanied by a hyperbolic hedgehog, the
  Saturn-ring configuration where the particle is surrounded by a $-1/2$ 
  disclination ring at the equator, and the surface-ring configuration.}
\label{fig.transition}
\end{figure}

Poulin\ {\em et al.\/}\ showed experimentally
that in inverted nematic emulsions, where 
surfactant-coated water droplets are dispersed in a nematic liquid crystal,
a director field configuration of dipolar symmetry occurs 
(see Fig.\ \ref{fig.transition}) \cite{itapdb:Poulin1997,itapdb:Poulin1998}.
The water droplet and its companion hyperbolic point defect form a
tightly bound object which we call dipole for short. A similar observation
at a nematic-isotropic interface was made by R.~B. Meyer much earlier
\cite{itapdb:Meyer1972}. Both the droplet
and the defect carry a topological charge $+1$, which ``add up'' to
the total charge 0 of the dipole \cite{itapdb:Kurik1988}. Theoretically
it was described with the help of ansatz functions that were
motivated by an electrostatic analog
\cite{itapdb:Poulin1997,itapdb:Lubensky1998}.
Terentjev {\em et al.\/} introduced the Saturn-ring configuration with
quadrupolar symmetry where a $-1/2$ disclination ring surrounds the
sphere at the equator (see Fig.\ \ref{fig.transition}).
It was investigated by both analytic and
numerical methods \cite{itapdb:Terentjev1995,itapdb:Kuksenok1996}.
By shrinking the disclination ring to the topologically equivalent
hyperbolic point defect, the Saturn ring can be continuously
transformed into the dipole configuration \cite{itapdb:Lubensky1998}.
With the help of an ansatz function that describes such a transformation
it was conjectured that for sufficently small particles the Saturn
ring should be more stable than the dipole \cite{itapdb:Lubensky1998}.
For a finite anchoring strength of the molecules at the surface a third
structure occurs, which we also illustrate in Fig.\ \ref{fig.transition}.
We call it the surface-ring configuration. Depending on the anchoring
strength $W$ there exists either a $-1/2$ disclination ring sitting
directly at the surface, or, for smaller $W$ \cite{itapdb:Kuksenok1996}, 
the director field is smooth everywhere, and
a ring of tangentially oriented molecules is located at the
 equator of the sphere. By means of a Monte-Carlo simulation
Ruhwandl and Terentjev showed that for sufficently small anchoring
strength $W$ the surface ring is the preferred configuration 
\cite{itapdb:Ruhwandl1997a}.

In this article we give a full account of the three director
configurations in Fig.\ \ref{fig.transition} by numerically minimizing
the Frank free energy supplemented by a magnetic-field and a surface
term \cite{itapdb:deGennes1993}. We go beyond the one-constant approximation,
generally used in the work cited above, and include the saddle-splay
term of the Frank free energy. Furthermore, the results of the
analytic approach based on the use of ansatz functions are checked.
In particular, we investigate the dipole configuration, which undergoes
a twist transition. Then, we show in detail that the transition from 
the dipole to the Saturn ring configuration can be either achieved by
decreasing the particle size or by applying a magnetic field. The role
of metastability is discussed. Finally, the surface ring is considered,
and the special role of the saddle-splay free energy for its
occurence is pointed out. Lower bounds for the surface-anchoring 
strength $W$ are given. All these results are presented in
Sec. \ref{sec.disc}. In Sec.\ \ref{sec.num} we define the
geometry of our problem, write down the reduced free energy, and
explain the numerical method to minimize it.

\begin{figure}
\centerline{\psfig{figure=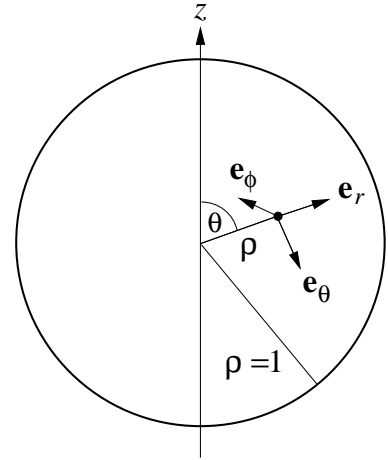,width=5cm}}

\vspace{0.3cm}

\caption[]{Illustration of the modified spherical coordinates
($\rho=1/r^{2},\theta$). At each point the local coordinate basis 
$({\bf e}_{r}, {\bf e}_{\theta}, {\bf e}_{\phi})$ of the standard 
spherical coordinate system is attached. The integration area is given by 
$0 \le \rho \le 1$.}
\label{fig.cosystem}
\end{figure}

\section{Geometry, Free Energy, and Numerical Method} \label{sec.num}

The director field around a spherical particle follows from the
minimization of the Frank free energy supplemented by a 
magnetic-field and a surface term. In this section we describe
our coordinate system, review the free energy, and give some
numerical details.

\subsection{Geometry} \label{subsec.geo}

The region outside the spherical particle with radius $a$ is
infinitely extended. We use a modified spherical coordinate system
with a radial coordinate $\rho = 1/r^{2}$ where $r$ is the distance
of a space point from the center of the particle measured in units of 
$a$. The exponent 2 is motivated by the far-field of the dipol
configuration \cite{itapdb:Poulin1997,itapdb:Lubensky1998}.
Such a transformation has two advantages. The
exterior of the particle is mapped into a finite region, {\em i.e.},
the interior of the unit sphere ($\rho \le 1$). 
Furthermore, equally spaced grid
points along the coordinate $\rho$ result in a mesh size in
real space which is small close to the surface of the particle. In
this area the director field is strongly varying, and hence a good 
resolution for the numerical calculation is needed. On the other hand,
the mesh size is large far away from the sphere where the director
field is nearly homogeneous. Since our system is axially symmetric,
the director field only depends on $\rho$ and
the polar angle $\theta$ as illustrated in Fig.\ \ref{fig.cosystem}. 
The symmetry axis corresponds to the $z$ axis. At each point $(\rho,\theta)$
we attach the local coordinate basis 
$({\bf e}_{r}, {\bf e}_{\theta}, {\bf e}_{\phi})$ of the standard 
spherical coordinate system and express the director in this basis:
\begin{equation}
{\bf n}(\rho,\theta) = n_{r}(\rho,\theta) {\bf e}_{r} +
 n_{\theta}(\rho,\theta) {\bf e}_{\theta} +
 n_{\phi}(\rho,\theta) {\bf e}_{\phi} \enspace.
\label{2.1}
\end{equation}
Since the director is a unit vector, we write
\begin{equation}
n_{r} = \cos \Theta \enspace, \enspace
n_{\theta} = \sin \Theta \cos \Phi \enspace, \enspace
n_{\phi} = \sin \Theta \sin \Phi \enspace,
\label{2.2}
\end{equation}
where $\Theta(\rho,\theta)$ and $\Phi(\rho,\theta)$ denote,
respectively, the tilt and the twist angle.
At the surface of the particle we allow the director to rotate
away from the preferred radial direction by introducing a surface
free energy (see next subsection). At infinity, {\em i.e.},
at $\rho = 0$ ${\bf n}$ 
always points along the $z$ axis. For completeness we note that
the differentials $dr$ and $d \rho$ are connected via
\begin{equation}
dr = - d \rho / (2 \rho^{3/2}) \enspace.
\label{2.3}
\end{equation}

\subsection{Free energy}

The free energy which we will minimize consists of bulk and surface
terms:
\begin{equation}
F_{\bf n} = \int d^{3}x (f_{el} + f_{24} + f_{H}) + \int dS f_{S} \enspace
\label{2.4}
\end{equation}
with the free energy densities:
\begin{mathletters}
\begin{eqnarray}
f_{el} & = & \frac{1}{2} \{ K_{1} (\bbox{\nabla} \cdot {\bf n})^{2}
  + K_{2} ({\bf n} \cdot \bbox{\nabla} \times {\bf n})^{2} \nonumber \\
 & &  + K_{3} [{\bf n} \times (\bbox{\nabla} \times {\bf n}) ]^{2} \}
\label{2.5a}
\end{eqnarray}
\begin{equation}
f_{24} = - \frac{1}{2} K_{24} \bbox{\nabla} \cdot 
[{\bf n} (\bbox{\nabla} \cdot {\bf n} ) + {\bf n} \times 
(\bbox{\nabla} \times {\bf n})]
\label{2.5b}
\end{equation}
\begin{equation}
f_{H} = -\frac{1}{2} \Delta \chi [({\bf n} \cdot {\bf H})^{2}
 - H^{2} ]
\label{2.5c}
\end{equation}
\begin{equation}
f_{S} = \frac{1}{2} W [ 1 - ({\bf n} \cdot \hat{\bbox{\nu}})^{2}]
\enspace.
\label{2.5d}
\end{equation}
\end{mathletters}
The Oseen-Z\"ocher-Frank free energy density $f_{el}+f_{24}$ describes
elastic distortions of the director field ${\bf n}({\bf r})$,
where $K_{1}$, $K_{2}$, $K_{3}$, and $K_{24}$ denote, 
respectively, the splay, twist, bend, and saddle-splay elastic constants.
The saddle-splay term is a pure divergence; it, therefore,
can be transformed into integrals over all surfaces of the system.
A Cauchy relation for $K_{24}$ follows from the Maier-Saupe molecular 
approach \cite{itapdb:Nehring1971}:
\begin{equation}
\label{2.6}
K_{24} = (K_{11} + K_{22}) / 2 \enspace.
\end{equation}
(There is also the possibility of another surface term with
a free energy density 
$K_{13} \bbox{\nabla} \cdot ({\bf n} \bbox{\nabla}\cdot {\bf n})$,
which we will not consider in this paper.)
Eq. (\ref{2.5c}) couples the director to an external magnetic field
${\bf H}$, where $\Delta \chi = \chi_{\|} - \chi_{\perp}$ stands for the
magnetic anisotropy.
The symbols $\chi_{\|}$ and $\chi_{\perp}$ denote the
magnetic susceptibilities for a magnetic field applied, respectively,
parallel or perpendicular to the director. 
In this paper we consider a positive $\Delta \chi$
that favors an alignment of the director ${\bf n}$ parallel to ${\bf H}$.
Since we calculate the magnetic free energy of the infinitely extended 
region around the sphere,
we use the magnetic free energy of a completely aligned
director field as a reference point to avoid infinities. As a result
the term $-\Delta \chi H^{2} /2$ in Eq. (\ref{2.5c}) occurs.
Finally, we employ the surface free energy of Rapini-Papoular to take
into account the interaction of the director with the boundaries.
In Eq. (\ref{2.5d}) the unit vector $\hat{\bbox{\nu}}$ denotes some
preferred orientation of the director at the surface,
and $W$ is the coupling constant. It varies
in the range $10^{-4} - 1$ erg/cm$^{2}$ as reviewed by Blinov 
{\em et al.\/} \cite{itapdb:Blinov1989a}. 
However, the authors do not specify $W$ for the
interface of water and the liquid-crystalline phase of 5CB
in the presence of the surfactant sodium dodecyl sulfate,
which was used in the experiment by Poulin {\em et al.\/} 
\cite{itapdb:Poulin1997,itapdb:Poulin1998}.
In subsection \ref{subsec.finite} we will give a lower bound for $W$
for such an interface.

For the numerical minimization a reduced form of the free
energy of Eq.\ (\ref{2.4}) is used. We introduce the energy unit 
$\pi K_{3} a$ and refer all lengths to the radius $a$ of the 
spherical particle. Furthermore, we employ the modified
spherical coordinates ($\rho,\theta$), take into account the axial 
symmetry of our system, and, finally, arrive at the reduced free energy 
\begin{eqnarray}
\label{2.7}
\overline{F}_{\bf n} & = & F_{\bf n} / (\pi K_{3} a) \\
\nonumber
 & = &\int_{\rho \le 1} 
d \theta d \rho \frac{\sin \theta}{\rho^{5/2}}
(\overline{f}_{el} + \overline{f}_{24} + \overline{f}_{H}) + 
\int_{\rho = 1} d \theta \sin \theta 
\overline{f}_{S} \enspace,
\end{eqnarray}
where
\begin{mathletters}
\begin{eqnarray}
\label{2.8a}
\overline{f}_{el} & = & \frac{1}{2} \{ \overline{K}_{1} 
 (\bbox{\nabla} \cdot {\bf n})^{2}
 + \overline{K}_{2} ({\bf n} \cdot \bbox{\nabla} \times {\bf n})^{2}
 \nonumber \\
 & & + [{\bf n} \times (\bbox{\nabla} \times {\bf n}) ]^{2} \}
\end{eqnarray}
\begin{equation}
\label{2.8b}
\overline{f}_{24} = - \frac{1}{2} \overline{K}_{24} \bbox{\nabla} \cdot 
[{\bf n} (\bbox{\nabla} \cdot {\bf n} ) + {\bf n} \times 
(\bbox{\nabla} \times {\bf n})]
\end{equation}
\begin{equation}
\label{2.8c}
\overline{f}_{H} = -\frac{1}{2} \xi_{H}^{-2} [(n_{r} \cos \theta - 
n_{\theta} \sin \theta)^{2} - 1]
\end{equation}
\begin{equation}
\label{2.8d}
\overline{f}_{S} = \xi_{S}^{-1} (1-n_{r}^{2}) \enspace.
\end{equation}
\end{mathletters}
In Eqs.\ (\ref{2.8a}) and (\ref{2.8b}) the coefficients
$\overline{K}_{1} = K_{1} / K_{3}$, $\overline{K}_{2} = K_{2} / K_{3}$,
and $\overline{K}_{24} = K_{24} / K_{3}$ denote, respectively, the
reduced splay, twist, and saddle-splay elastic constants. The single
contributions to the Frank free energy density in our modified spherical
coordinates are rather lengthy, and we refer the reader to 
Appendix\ \ref{app.A} for the detailed form. 
We always apply the magnetic field 
${\bf H}$ along the symmetry axis of our system, which coincides 
with the $z$ axis (see Fig. \ref{fig.cosystem}). Inserting
${\bf H} = H {\bf e}_{z}$ with
${\bf e}_{z} = \cos \theta {\bf e}_{r} - \sin \theta {\bf e}_{\theta}$
into Eq. (\ref{2.5c}) we obtain the magnetic free energy density
of Eq. (\ref{2.8c}). The strength of the field is given via the reduced
magnetic coherence length:
\begin{equation}
\label{2.9}
\xi_{H} = \sqrt{K_{3}/(\Delta \chi H^{2})} / a \enspace.
\end{equation}
It indicates the distance in the bulk which is needed to orient the
director along the applied field when, {\em e.g.}, through boundary
conditions a different preferred orientation of the molecules exists
\cite{itapdb:deGennes1993}.
The length tends to infinity for $H \rightarrow 0$.
The surface term of Eq. (\ref{2.8d}) follows from Eq. (\ref{2.5d})
by choosing $\hat{\bbox{\nu}} = {\bf e}_{r}$, {\em i.e.}, a preferred
radial anchoring of the molecules at the surface of the suspended particle.
The reduced extrapolation length \cite{itapdb:deGennes1993}
\begin{equation}
\label{2.10}
\xi_{S} = K_{3} / (Wa)
\end{equation}
signifies the strength of the anchoring. It compares the Frank free
energy of the bulk, which is proportional to $K_{3} a$, to the 
surface energy, which scales with $Wa^{2}$.
At {\em strong\/} anchoring, {\em i.e.}, for $\xi_{S} \ll 1$ the
energy to rotate the director away from its preferred direction 
$\hat{\bbox{\nu}}$ at the whole surface would be much larger 
than the bulk energy. Therefore, it is preferable for the system when
the director points along $\hat{\bbox{\nu}}$. However, ${\bf n}$ can
deviate from $\hat{\bbox{\nu}}$ in the area $\xi_{S} a^{2}$.
This will explain the result in subsection \ref{subsec.finite} where
we show that the surface ring already appears at strong surface coupling.
{\em Rigid\/} anchoring is realized for $\xi_{S} \rightarrow 0$, and
$\xi_{S} \gg 1$ means {\em weak\/} anchoring, where the influence of
the surface is minor.

To calculate the free energy $\overline{F}_{\bf n}$ numerically we transform 
the saddle-splay term into a surface integral:
\begin{equation}
\label{2.11}
\overline{F}_{24} = \overline{K}_{24} \int_{\rho=1} d\theta
\sin \theta {\bf e}_{r} \cdot [{\bf n} (\bbox{\nabla} \cdot {\bf n})
 + {\bf n} \times (\bbox{\nabla} \times {\bf n}) ] \enspace.
\end{equation}
The detailed form follows from Appendix\ \ref{app.A}.
In performing Gauss' theorem we only include the surface of the sphere.
There is also a contribution from the surface of the core of a
possible disclination ring. However, since we can only describe the energy
of the core in an approximate way (see next paragraph), we will skip
this term. For rigid homeotropic anchoring ($n_{r}=1$, 
$n_{\theta} = n_{\phi}=0$) the free energy of the saddle-splay term is
\begin{equation}
\label{2.12}
\overline{F}_{24} = 4 \overline{K}_{24} \enspace.
\end{equation}
To arrive at Eq.\ (\ref{2.12})
we used $\partial n_{r} / \partial \rho = 0$, which is valid because
of the normalization of the director.

The free energy of a $\pm\frac{1}{2}$ disclination ring is taken
into account by the line energy $F_{d}$ of a $\pm\frac{1}{2}$ disclination. 
In the one-constant approximation 
($K_{1}=K_{2}=K_{3}=K$) and in reduced units it reads 
\cite{itapdb:Kleman1983}
\begin{equation}
\label{2.13}
\overline{F}_{d} = F_{d}/K = \frac{\pi}{4} \left(\frac{1}{2} + 
\ln \frac{R}{r_{c}} \right) \enspace.
\end{equation}
The first term denotes the line energy of the core with a core radius 
$r_{c}$ which in absolute lengths is of the order
of $10\,\text{nm}$. The second term stands for the elastic energy
of the disclination line where $R$ is its radial extent [see Fig.\ 
\ref{fig.ringcoord}b)].
In the general case ($K_{1} \ne K_{2} \ne K_{3}$) an analytic
expression for the elastic energy does not exist. We only use
a rough approximation for the core energy $F_{c}$ by averaging over 
the Frank constants:
\begin{equation}
\label{2.14}
\overline{F}_{c} = F_{c} / K_{3} = \frac{\pi}{8} (\overline{K}_{1} 
+ \overline{K}_{2} +1) / 3 \enspace.
\end{equation}
A more quantitative description of the free energy of a disclination
has to start from the Landau-Ginzburg-de Gennes free energy with
the full alignment tensor ${\bf Q}$ \cite{itapdb:Schopohl1987,itapdb:Penzenstadler1989}.

We finish this subsection with an important remark about length
scales. In the free energy of Eq.\ (\ref{2.7}) all lengths are given
relative to the particle radius $a$, which appears in the 
absolute energy unit $\pi K_{3} a$ only. This would suggest that the 
director configuration does not depend on the particle size. However, 
with the core radius $r_{c}$ a second length scale is introduced,
which in absolute lengths is always of the order of $10\,\text{nm}$
\cite{itapdb:Kleman1983}.
On the other hand, in the next subsection we explain that
$r_{c}$ in units of $a$ is a parameter of our numerical
method whose lower bound is determined by the mesh size of the grid.
In the discussions of Sec.\ \ref{sec.disc} we want to make a
connection to experiments.
Therefore, we do not give this dimensionless $r_{c}$, but use an
absolute core size of $10\,\text{nm}$ to calculate the corresponding
particle radius $a$:
\begin{equation}
\label{2.14b}
a = 10\,\text{nm}/ r_{c} \enspace.
\end{equation}
In subsection \ref{subsubsec.size} we will study how the configuration
around a spherical particle depends on $a$.

\subsection{Minimization and numerical details}
The fields of the tilt [$\Theta(\rho,\theta)$] and the twist 
[$\Phi(\rho,\theta)$] angle follow from a minimization of the reduced
free energy $\overline{F}_{\bf n}[\Theta,\Phi]$ of Eq.\ (\ref{2.7}). 
The corresponding  Euler-Lagrange equations are equivalent to the functional 
derivatives of $\overline{F}_{\bf n}[\Theta,\Phi]$:
\begin{eqnarray}
\label{2.15a}
\frac{\delta \overline{F}_{\bf n}}{\delta \Theta} & = &
\frac{\delta \overline{F}_{\bf n}}{\delta n_{i}} 
\frac{\partial n_{i}}{\partial \Theta} = 0 \\
\label{2.15b}
\frac{\delta \overline{F}_{\bf n}}{\delta \Phi}  & = &
\frac{\delta \overline{F}_{\bf n}}{\delta n_{i}} 
\frac{\partial n_{i}}{\partial \Phi} = 0 \enspace,
\end{eqnarray}
where $i$ stands for $r$, $\theta$, and $\phi$ and Einstein's
summation convention over repeated indices is used.
We have employed a chain rule to arrive at the Euler-Lagrange
equations for $\Theta(\rho,\theta)$ and $\Phi(\rho,\theta)$, which is 
generally valid for functional derivatives as shown in Appendix\ \ref{app.B}.
Performing the variation of the free energy $\overline{F}_{\bf n}$ we arrive
at the Euler-Lagrange equations in the bulk,
\begin{eqnarray}
\label{2.16a}
\left[\left( \frac{\partial}{\partial n_{i}} - 
\frac{\partial}{\partial x_{j}} \frac{\partial}{\partial n_{i,j}} \right)
\frac{\sin \theta}{\rho^{5/2}} (\overline{f}_{el} + \overline{f}_{H}) \right]
\frac{\partial n_{i}}{\partial \Theta} & = & 0\\
\label{2.16b}
\left[ \left( \frac{\partial}{\partial n_{i}} - 
\frac{\partial}{\partial x_{j}} \frac{\partial}{\partial n_{i,j}} \right)
\frac{\sin \theta}{\rho^{5/2}} (\overline{f}_{el} + \overline{f}_{H}) \right]
\frac{\partial n_{i}}{\partial \Phi} & = & 0 \enspace,
\end{eqnarray}
\begin{figure}
\centerline{\psfig{figure=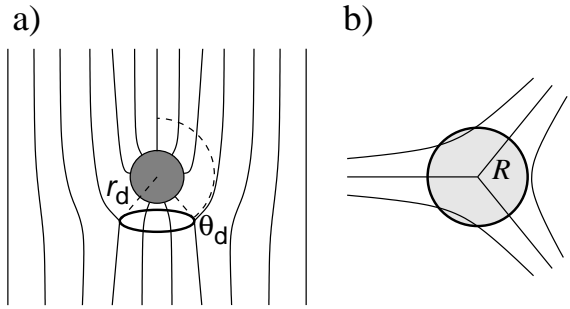,width=7.5cm}}

\vspace{0.5cm}

\caption[]{a) Coordinates ($r_{d},\theta_{d}$) for a -1/2 disclination
  ring with a general position around the spherical particle; b)
  Cross section of the $-1/2$ disclination ring with a radial
  extension $R$.}
\label{fig.ringcoord}
\end{figure}
\noindent
and at the surface,
\begin{eqnarray}
\label{2.17a}
\left(\frac{\partial \overline{f}_{S}}{\partial n_{i}} + 
\frac{\partial}{\partial n_{i,\rho}} (\overline{f}_{el} + \overline{f}_{24})
\right) \frac{\partial n_{i}}{\partial \Theta} & = & 0\\
\label{2.17b}
\left(\frac{\partial \overline{f}_{S}}{\partial n_{i}} + 
\frac{\partial}{\partial n_{i,\rho}} (\overline{f}_{el} + \overline{f}_{24})
\right) \frac{\partial n_{i}}{\partial \Phi} & = & 0 \enspace,
\end{eqnarray}
where $n_{i,j}$ stands for $\partial n_{i} / \partial x_{j}$.
The Euler-Lagrange equations are calculated with the help of the 
algebraic program Maple and are then imported into a Fortran program where
they are usually solved on a rectangular $33 \times 129$ grid in the 
$(\rho,\theta)$ space. In Eq.~(33) of ref.~\cite{itapdb:Lubensky1998} 
an analytical form of
the director configuration of a $-\frac{1}{2}$ disclination ring
around a spherical particle is given. The position of this ring is
determined by the radial ($r_{d}$) and the angular ($\theta_{d}$)
coordinates [see Fig.~\ref{fig.ringcoord}a)]. We take this director 
configuration and let it relax via the standard 
Newton-Gauss-Seidel method \cite{itapdb:Press1992}.
In the dipole configuration ($\theta_{d} = 0, \pi$) the hyperbolic
point defect moves along the $z$ axis in its local minimum when the
numerical relaxation is performed. However,
a disclination ring ($\theta_{d} \ne 0, \pi$)
basically stays at the positon where we place it. We use this fact
to investigate the free energy as a function of $r_{d}$ and $\theta_{d}$
which gives an instructive insight into potential barriers for a 
transition between the dipole and the Saturn-ring configuration.

The free energy $\overline{F}_{\bf n}$ of the director field follows
from a numerical integration. This procedure assigns some energy to
the disclination ring which certainly is not correct.
To obtain a more accurate value for the total free energy $\overline{F}$
we use the formula
\begin{equation}
\label{2.18}
\overline{F} = \overline{F}_{\bf n} - \left. \overline{F}_{\bf n}
\right|_{\text{torus}} + 
\overline{F}_{c/d} \times 2\pi r_{d} \sin\theta_{d} \enspace.
\end{equation}
The quantity $\left. \overline{F}_{\bf n} \right|_{\text{torus}}$ denotes the
numerically calculated free energy of a toroidal region of cross
section $\pi R^{2}$ around the disclination ring 
[see Fig.\ \ref{fig.ringcoord}b)]. Its volume is 
$\pi R^{2} \times 2\pi r_{d} \sin\theta_{d}$, where the coordinates
$(r_{d},\theta_{d})$ of the ring are determined by searching for the
maximum of the local free energy density
$\overline{f}_{el}$. 
The value $\left. \overline{F}_{\bf n} \right|_{\text{torus}}$ is
replaced by the last term on the
right-hand side of Eq.\ (\ref{2.18}), which
provides the correct free energy according to Eqs.\
(\ref{2.13}) or (\ref{2.14}). To find out how large the cross section
$\pi R^{2}$ of the cut torus has to be, we employed the last formula
and the line energy of Eq.\ (\ref{2.13}) for constant $r_{c}$ and
varying $R$. To be consistent, $\overline{F}$ should not depend on
$R$. Within an error of less than 1\ \% this is the case if
$\pi R^{2}$ is equal or larger than
$3 \Delta \rho \Delta \theta / 2$ where $\Delta \rho$ and 
$\Delta \theta$ are the lattice constants of our grid. 
To study the transition between the dipole and the Saturn ring
as a function of the particle size we choose 
$\pi R^{2} = 25 \Delta \rho \Delta \theta /2$, employ Eq.\ (\ref{2.13})
for different $r_{c}$, and calculate $a$ from Eq.\ (\ref{2.14b}).
In all other cases we set $R=r_{c}$, determine $r_{c}$ from
$\pi r^{2}_{c} = 3 \Delta \rho \Delta \theta /2$, and use the core energy of
Eq.\ (\ref{2.14}).

The radial extension of the core of a point defect is also of the
order of 10\,nm \cite{itapdb:Lavrentovich1986}, and its free energy
is approximated by $K_{3}\times 10\,\mathrm{nm}$. As we show in the 
following section the free energy of the dipole amounts to around
$10\pi K_{3}a$. 
Since we consider particle radii larger than $100\,\mathrm{nm}$,
the contribution from the energy of the point defect is
always smaller than 1 \%. This is beyond our numerical accuracy, and 
therefore no energetical correction for the point defect was included.

The discussion in the following section always uses the one-constant
approximation or the nematic liquid crystal pentylcyanobiphenyl (5CB)
with the bend elastic constant $K_{3} = 0.53 \times 10^{-6}\,\mathrm{dyn}$
and the reduced splay and twist elastic constants $\overline{K}_{1} = 0.79$
and $\overline{K}_{2} = 0.43$, respectively.

\section{Results and Discussion} \label{sec.disc}
In this section we present the results of our numerical
investigation. We first address a twist transition in the
dipole configuration. Then we discuss a transition from the dipole
to the Saturn ring which is induced either by decreasing the particle
size or by applying a magnetic field. Finally, we illustrate that 
the surface-ring configuration appears when the surface-anchoring
strength is lowered.

\subsection{Twist transition of the dipole configuration}

\begin{figure}

\vspace{-1.1cm}

\centerline{\psfig{figure=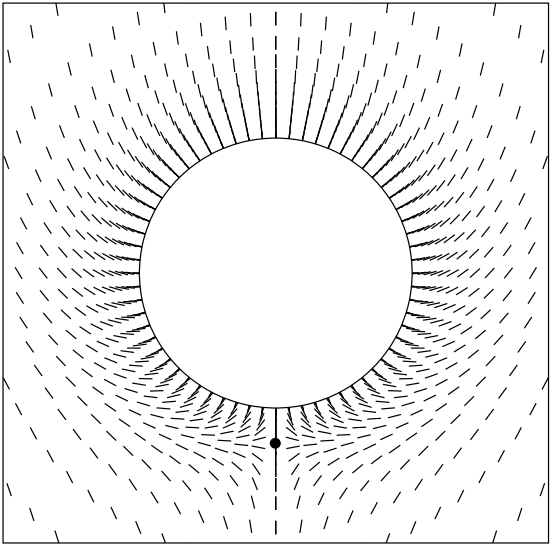,width=7.5cm}}

\vspace{0.5cm}

\caption[]{The numerically determined director field of the dipole 
configuration for the one-constant approximation, zero magnetic
field, and rigid surface anchoring. The location of the hyperbolic hedgehog
is indicated by a dot. The directors lie in the drawing plane.}
\label{fig.dirfield}
\end{figure}

In Fig.\ \ref{fig.dirfield} we plot the director field of the dipole
configuration for the one-constant approximation. A magnetic field is
not applied and the directors are rigidly anchored at the surface.
The dot indicates the location of the hyperbolic hedgehog. 
For its distance $r_{d}$ from the center
of the sphere we find $r_{d} = 1.26 \pm 0.02$, where the mesh size of
the grid determines the uncertainty in $r_{d}$. Our result is in
excellent agreement with Ref. \cite{itapdb:Lubensky1998}. In this article
the dipole was described via an ansatz function. However, 
Ruhwandl and Terentjev using a Monte-Carlo minimization report a
somewhat smaller value for $r_{d}$ \cite{itapdb:Ruhwandl1997}.

\begin{figure}

\vspace{-.5cm}

\centerline{\psfig{figure=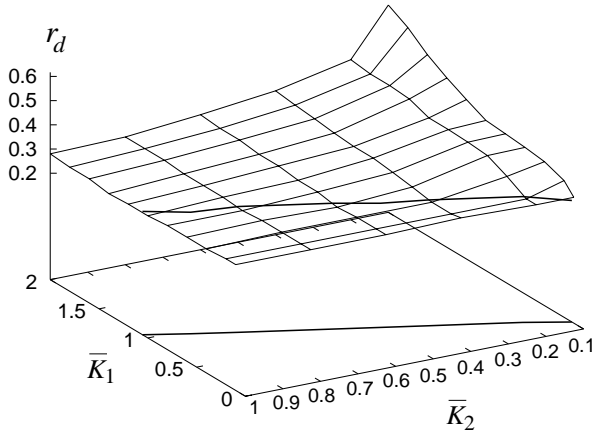,width=10cm}}
\caption[]{The distance $r_{d}$ of the hyperbolic hedgehog from the
center of the sphere as a function of the reduced splay 
($\overline{K}_{1}$) and twist ($\overline{K}_{2}$) constants.}
\label{fig.dis}
\end{figure}

\begin{figure}
\centerline{\psfig{figure=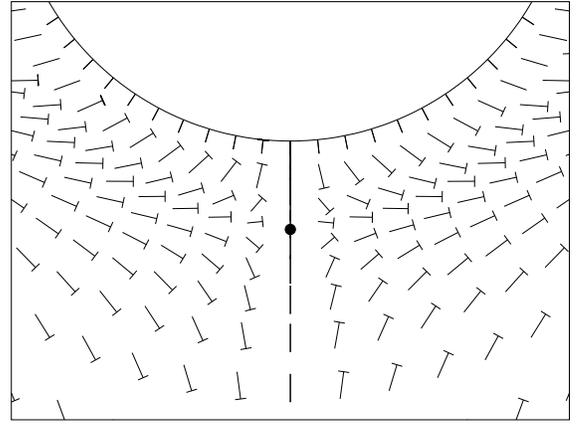,width=7.5cm}}

\vspace{.5cm}

\caption[]{Nail picture of a close-up of the twisted dipole configuration. 
Around the hyperbolic hedgehog the directors are tilted out of the 
drawing plane. The length of the nail is proportional to the
projection of the director on the drawing plane. The head of the nail
is below the plane.}
\label{fig.nail}
\end{figure}

Fig.\ \ref{fig.dis} presents the distance $r_{d}$ as a function
of the reduced splay ($\overline{K}_{1}$) and twist ($\overline{K}_{1}$)
constants. In front\linebreak
\noindent
of the thick line $r_{d}$ is basically
constant. Beyond the line $r_{d}$ starts to grow which indicates a
structural change in the director field illustrated in the nail picture of
Fig. \ref{fig.nail}. Around the hyperbolic hedgehog the directors
develop a non-zero azimuthal component $n_{\phi}$ introducing a twist
into the dipole. It should be visible under a polarizing
microscope when the dipole is viewed along its symmetry axis. 

\begin{figure}
\centerline{\psfig{figure=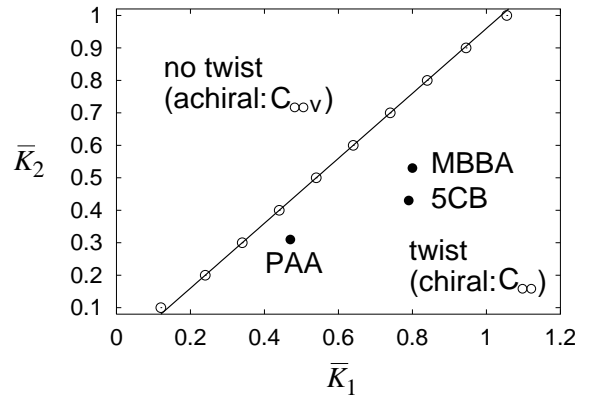,width=7.5cm}}

\vspace{0.2cm}

\caption[]{Phase diagram of the twist transition as a function of
the reduced splay ($\overline{K}_{1}$) and twist ($\overline{K}_{2}$) 
constants. A full explanation is given in the text.}
\label{fig.twist}
\end{figure}

In Fig. \ref{fig.twist} we draw a phase diagram for the twist
transition. As expected, it occurs when $\overline{K}_{1}$ increases
or when $\overline{K}_{2}$ decreases, {\em i.e.}, when a twist
deformation costs less energy than a splay distortion.
The open circles are nu-\linebreak
\noindent
merical results for the transition line which
can be well fitted by the straight line 
$\overline{K}_{2} \approx \overline{K}_{1} - 0.04$. Interestingly,
the small offset $0.04$ means that $K_{3}$ does not play an important
role. Typical calamatic liquid crystals like MBBA, 5CB, and PAA should
show the twisted dipole configuration.

Since the twist transition breaks the mirror symmetry of the dipole,
which then becomes a chiral object, we describe it by a Landau expansion
of the free energy:
\begin{equation}
\label{3.1}
F = F_{0} + a(\overline{K}_{1},\overline{K}_{2}) [n_{\phi}^{\text{max}}]^{2}
+ c [n_{\phi}^{\text{max}}]^{4} \enspace.
\end{equation}
With the maximum azimuthal component $n_{\phi}^{\text{max}}$ we have
introduced a simple order parameter. For symmetry reasons only
even powers of $n_{\phi}^{\text{max}}$ are allowed.
The phase transition line is determined by
$a(\overline{K}_{1},\overline{K}_{2}) = 0$. According to Eq.\ (\ref{3.1})
we expect a power-law dependence of the order parameter with the exponent
$1/2$ in the twist region\linebreak
\begin{figure}
\centerline{\psfig{figure=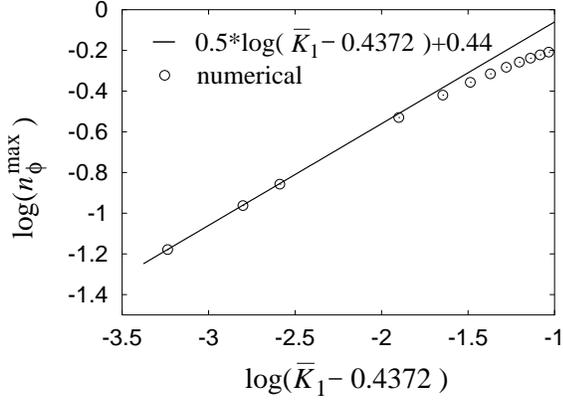,width=7.5cm}}

\vspace{0.2cm}

\caption[]{Log-log plot of the order parameter $n_{\phi}^{\text{max}}$
versus $\overline{K}_{1}$ close to the twist transition for 
$\overline{K}_{2} = 0.4$; $\circ$ \dots numerical values, $-$ \dots fit 
by a straight line.}
\label{fig.orderp}
\end{figure}
\noindent
close to the phase transition. To test this
idea we choose a constant $\overline{K}_{2}$ and determine
$n_{\phi}^{\text{max}}$ for varying $\overline{K}_{1}$. As the log-log
plot in Fig.\ \ref{fig.orderp} illustrates, when approching the phase
transition, the order parameter obeys the expected power law:
\begin{equation}
\label{3.2}
n_{\phi}^{\text{max}} \sim (\overline{K}_{1} - 0.4372)^{1/2}
\enspace,\qquad \overline{K}_{2} = 0.4 \enspace.
\end{equation}

\subsection{Dipole versus Saturn ring} \label{subsec.dipversat}

There are two possibilities to induce a transition from the dipole
to the Saturn-ring configuration; either by reducing the particle
size or by applying, {\em e.g.}, a magnetic field. We always assume
rigid anchoring in this subsection, set $K_{24} = 0$, and start with
the first point.

\subsubsection{Effect of particle size} \label{subsubsec.size}

\begin{figure}
\centerline{\psfig{figure=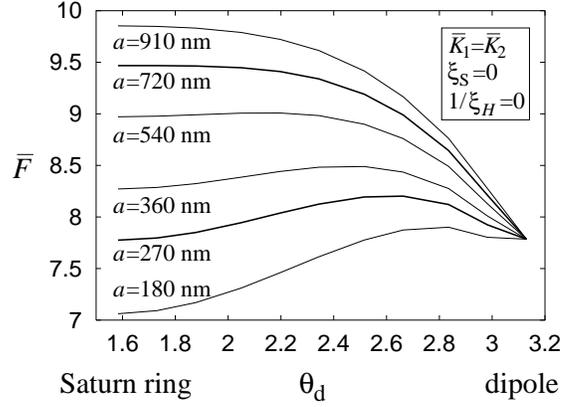,width=7.5cm}}

\vspace{0.2cm}

\caption[]{The reduced free energy $\overline{F}$ as a function of
the angular coordinate $\theta_{d}$. The parameter of the curves is 
the particle size $a$. Further parameters are indicated in the inset.}
\label{fig.size}
\end{figure}

In Fig. \ref{fig.size} we plot the reduced free energy $\overline{F}$
as a function of the angular coordinate $\theta_{d}$ of the disclination ring.
For constant $\theta_{d}$ the free energy $\overline{F}$ was chosen as
the minimum over the radial coordinate $r_{d}$.
The particle radius $a$ is the parameter of the curves and the one-constant
approximation is employed. Recall that $\theta_{d} = \pi/2$ and 
$\theta_{d} = \pi$ correspond, respectively, to the Saturn-ring or the
dipole configuration. Clearly, for small particle sizes 
($a = 180\,\text{nm}$) the
Saturn ring is the absolutely stable configuration and the dipole
enjoys some metastability. However, thermal fluctuations cannot induce
a transition to the dipole since the potential barriers are much
higher than the thermal energy $k_{\text{B}}T$. {\em E.g.\/}, a barrier of
$0.1 \pi K_{3} a$ corresponds to $1000\,k_{\text{B}}T$ 
($T = 300\,\text{K}$, $a=1\,\mu\text{m}$).
At $a \approx 270 \, \text{nm}$ the dipole assumes the
global minimum of the free energy, and finally the Saturn ring becomes
absolutely unstable at $a \approx 720\,\text{nm}$. The scenario agrees
with the findings of Ref. \cite{itapdb:Lubensky1998}. Furthermore, we stress
that\linebreak
\noindent
the particle sizes were calculated from Eq.\ (\ref{2.14b})
with the choice of $10\,\text{nm}$ as the real core size, and that our
results depend on the line energy (\ref{2.13}) of the disclination.

\begin{figure}
\centerline{\psfig{figure=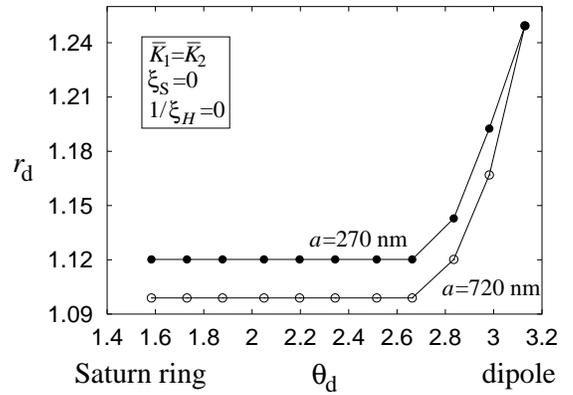,width=7.5cm}}

\vspace{0.2cm}

\caption[]{The preferred radial coordinate $r_{d}$ of the disclination
  ring in units of $a$ as a function of $\theta_{d}$ for two particle sizes. 
  Further parameters are indicated in the inset.}
\label{fig.radius}
\end{figure}

The preferred radial coordinate $r_{d}$ of the disclination ring 
as a function of $\theta_{d}$ is presented in Fig. \ref{fig.radius}. 
As long as the ring is open $r_{d}$ does not depend on $\theta_{d}$
within an error of $\pm 0.01$.
Only in the region where it closes down to the hyperbolic hedgehog
does $r_{d}$ increase sharply. The figure also illustrates
that the ring sits closer to larger particles. The radial position of
$r_{d}=1.10$ for $720\text{nm}$ particles agrees very well with Refs.
\cite{itapdb:Lubensky1998} and \cite{itapdb:Ruhwandl1997}.

\subsubsection{Effect of a magnetic field}

A magnetic field applied along the symmetry axis of the dipole can
induce a transition to the Saturn-ring configuration. This can be
understood from a simple back-of-the-envelope calculation. Let us
consider high magnetic fields, {\em i.e.\/}, magnetic coherence lengths
much smaller than the particle size $a$, which in our reduced\linebreak
\begin{figure}
\centerline{\psfig{figure=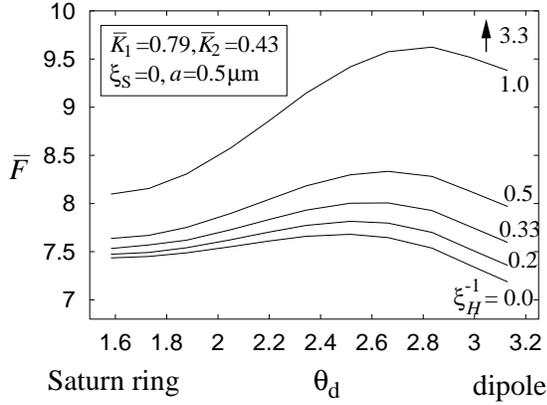,width=7.5cm}}

\vspace{0.2cm}

\caption[]{The reduced free energy $\overline{F}$ as a function
of the angular coordinate $\theta_{d}$. The parameter of the curves
is the inverse reduced magnetic coherence length $\xi_{H}^{-1}$.
Further parameters are indicated in the inset.}
\label{fig.magH}
\end{figure}
\noindent
units means:
$\xi_{H} \ll 1$. The directors are basically aligned along the magnetic field.
In the dipole configuration the director field close to the hyperbolic
hedgehog cannot change its topology. The field lines are "compressed" 
along the $z$ direction, and high densities of the elastic and
magnetic free energies occur in a region of thickness $\xi_{H}$.
Since the field lines have to bend around the sphere the cross section
of the region, in units of $a^{2}$, is of the order of 1, and its
volume is proportional to $\xi_{H}$. The Frank free energy
density is of the order of $1/\xi_{H}^{2}$, and
therefore the elastic free energy, in reduced units, scales with 
$1/\xi_{H}$. The same holds for the magnetic free energy. 
In the case of the Saturn-ring
configuration high free energy densities occur in a toroidal region
of cross section $\propto \xi_{H}^{2}$ around the disclination
ring. Hence, the volume scales with $\xi_{H}^{2}$ and the total
free energy is of the order of 1, {\em i.e.}, a factor $1/\xi_{H}$
smaller than for the dipole.

\begin{figure}

\vspace{-0.5cm}

\centerline{\psfig{figure=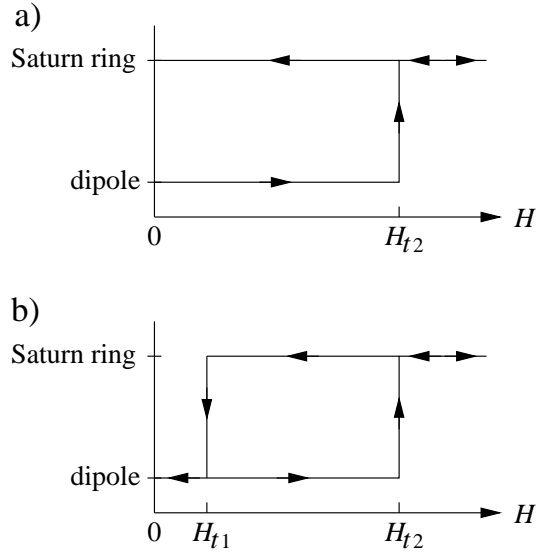,width=7cm}}

\vspace{0.3cm}

\caption[]{a) The Saturn ring is metastable at $H=0$. The dipole can be
transformed into the Saturn ring by increasing the magnetic field $H$
beyond $H_{t2}$ where the dipole looses its metastability. Turning off the
field the Saturn ring stays. b) The Saturn ring is unstable at $H=0$. When
the magnetic field is decreased from values above $H_{t2}$, the Saturn
ring shrinks back to the dipole at $H_{t1}$ where the Saturn ring
looses its metastability. A hysteresis occurs.}
\label{fig.machineII}
\end{figure}

Fig.\ \ref{fig.magH} presents a calculation for a particle size of
$a=0.5\,\mu\text{m}$ and the liquid crystal 5CB. We plot the reduced free
energy as a function of $\theta_{d}$ for different magnetic field
strengths given in units of the inverse reduced coherence 
length $\xi_{H}^{-1}$.
Without a field ($\xi_{H}^{-1} = 0$) the dipole is the 
energetically preferred configuration. The Saturn ring shows
metastability. A thermally induced transition between both states
cannot happen because of the high potential barrier. At a field
strength $\xi_{H}^{-1} = 0.33$ the Saturn ring becomes the
stable configuration. However, there will be no transition until the
dipole looses its metastability at a field strength $\xi_{H}^{-1} =
3.3$, which is only indicated by an arrow in Fig. \ref{fig.magH}.
Once the system has changed to the Saturn ring it will stay there even
for zero magnetic field.
Fig.\ \ref{fig.machineII}a) schematically illustrates how a dipole
can be transformed into a Saturn ring with the help of a magnetic
field. If the Saturn ring is unstable at zero field, a hysteresis
occurs [see Fig.\ \ref{fig.machineII}b)]. Coming from high magnetic
fields the Saturn ring looses its metastability at $H_{t1}$ and a
transition back to the dipole takes place.
In Fig.\ \ref{fig.size} we showed that the second situa-\linebreak
\noindent
tion is realized
for particles larger than $720\,\text{nm}$.
We also performed calculations for a particle size of 
$1\mu\text{m}$ and the liquid crystal 5CB and still find the Saturn
ring to be metastable at zero field in contrast to the result of the
one-constant approximation.

\begin{figure}
\centerline{\psfig{figure=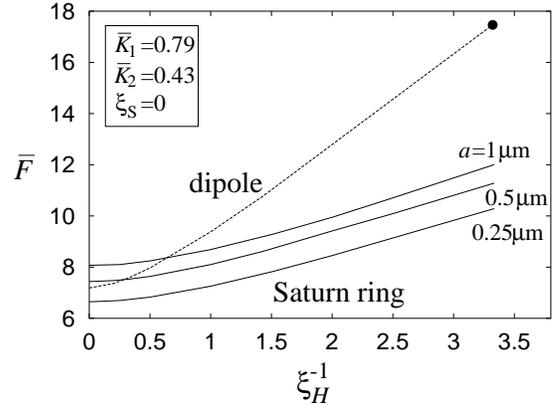,width=7.5cm}}

\vspace{0.2cm}

\caption[]{The reduced free energy $\overline{F}$ as a function of the
inverse coherence length $\xi_{H}^{-1}$ for the dipole and the Saturn
ring. The parameter of the curves is the particle size $a$. Further 
parameters are indicated in the inset.}
\label{fig.magHsi}
\end{figure}

Finally, in Fig. \ref{fig.magHsi} we plot the reduced free energy versus 
the applied magnetic field for different particle sizes. The energy of the 
dipole (dashed line) does not depend on $a$. However, the field
strength where the dipole be-\linebreak
\noindent
comes unstable should depend on the
particle size. The dot indicates this strength for $a=0.5\,\mu\text{m}$.
The energy curves of the Saturn ring are merely shifted by a constant
amount when the particle size $a$ is changed. For
$a=0.25 \, \mu\text{m}$ the Saturn ring is the stable configuration.
At the intersection points of full and dashed curves the absolute
stability changes from the dipole to the Saturn ring. These points
seem to occur at higher fields when the particle size is increased.
However, if we take into account that $\xi_{H}^{-1} \propto H$ is given 
relative to $a^{-1}$ there is not much variation in the absolute
field strength $H$ between the two particle sizes of 0.5 and 
$1\,\mu\text{m}$. Larger particles could not be investigated because
they would have required smaller mesh sizes of the grid.

\subsection{Influence of finite surface anchoring} \label{subsec.finite}

\begin{figure}
\centerline{\psfig{figure=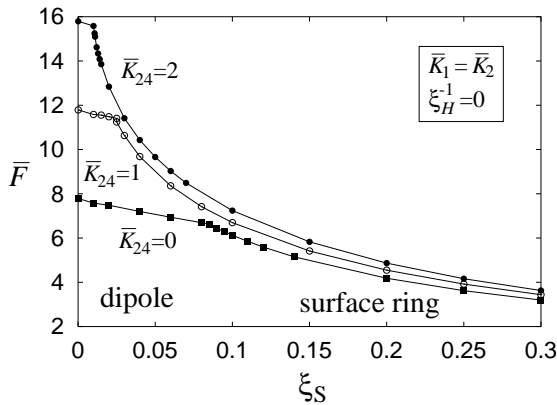,width=7.5cm}}

\vspace{0.2cm}

\caption[]{The minimum free energy $\overline{F}$ as a
  function of the surface extrapolation length $\xi_{S}$ for different
  $\overline{K}_{24}$. A first-order phase transition from the dipole
  to the surface ring occurs. Further parameters are indicated in
  the inset.}
\label{fig.surfaceK24}
\end{figure}

\begin{figure}
\centerline{\psfig{figure=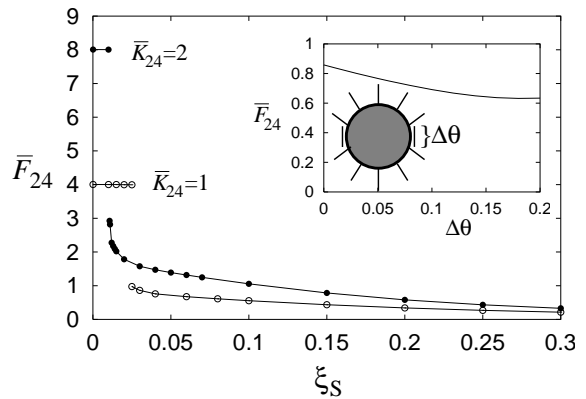,width=7.5cm}}

\vspace{0.2cm}

\caption[]{The saddle-splay free energy $\overline{F}_{24}$ as a
  function of $\xi_{S}$ for the same curves as in Fig.\
  \ref{fig.surfaceK24}. Inset: $\overline{F}_{24}$ versus the angular
  width of the surface ring calculated from the ansatz functions in
  Eqs.\ (\ref{3.3}) for $\overline{K}_{24}=1$.}
\label{fig.K24}
\end{figure}

In the last subsection we investigate the effect of finite anchoring
on the director field around the spherical particle. The saddle-splay
term with its reduced constant $\overline{K}_{24}$ is important
now. We always choose a zero magnetic field. In Fig.\ \ref{fig.surfaceK24}
we employ the one-constant approximation and plot
the free energy versus the surface extrapolation length $\xi_{S}$ for
different saddle-splay constants $\overline{K}_{24}$. Recall that
$\xi_{S}$ is inversely proportional to the surface constant $W$ 
[see Eq.\ (\ref{2.10})]. The straight lines belong to the dipole.
Then, for decreasing surface anchoring, there is a first-order
transition to the surface-ring structure. We never find the 
Saturn-ring to be the stable configuration although it enjoys some
metastability. For $\overline{K}_{24}=0$ the transition takes place at
$\xi_{S} \approx 0.085$. This value is somewhat smaller than the result
obtained by Ruhwandl and Terentjev \cite{itapdb:Ruhwandl1997}. 
One could wonder that the surface ring already occurs at such a strong
anchoring like $\xi_{S} \approx 0.085$ where any deviation from the
homeotropic anchoring costs a lot of energy. However, if 
$\Delta \theta$ is the angular width of the surface ring where
the director de-\linebreak
\noindent
viates from the homeotropic alignment 
(see inset of Fig.\ \ref{fig.K24}) then a simple energetical estimate
allows $\Delta \theta$ to be of the order of $\xi_{S}$.
It is interesting that the
transition point shifts to higher anchoring strengths, {\em i.e}, 
decreasing $\xi_{S}$
when $\overline{K}_{24}$ is increased. Obviously the saddle-splay term
favors the surface-ring configuration. To check this conclusion we
plot in Fig.\ \ref{fig.K24} the saddle-splay free energy $\overline{F}_{24}$
versus $\xi_{S}$. The horizontal lines belong to the dipole. They
correspond to the energy which one expects for a rigid homeotropic
anchoring at the surface of the sphere [see Eq.\ \ref{2.12})]. In
contrast, for the surface-ring configuration the saddle-splay energy
drops sharply. The surface ring around the equator of the sphere
introduces a ``saddle'' in the director field as illustrated in the
inset of Fig.\ \ref{fig.K24}. Such structures are known to be favored
by the saddle-splay term. We modeled the surface ring with an
angular width $\Delta \theta$ by the following director components:
\begin{mathletters}
\label{3.3}
\begin{equation}
n_{r} = \displaystyle - \tanh \left( 
\frac{\theta - \pi/2}{\Delta \theta}\right)
\end{equation}
\begin{equation}
n_{\theta} = \displaystyle -\left[ \cosh \left( 
\frac{\theta - \pi/2}{\Delta \theta}\right) \right]^{-1} \enspace,
\end{equation}
\end{mathletters}
where $\Delta \theta \ll \pi / 2$ to ensure that $n_{r} =1$ at 
$\theta = 0,\pi$, 
and calculated the saddle-splay energy versus $\Delta \theta$
by numerical integration. The result for $\overline{K}_{24} =1$ is
shown in the inset of Fig.\ \ref{fig.K24}. It fits very well to the
full numerical calculations and confirms again that a narrow
``saddle'' around the equator can considerably reduce the saddle-splay
energy.

\begin{figure}
\centerline{\psfig{figure=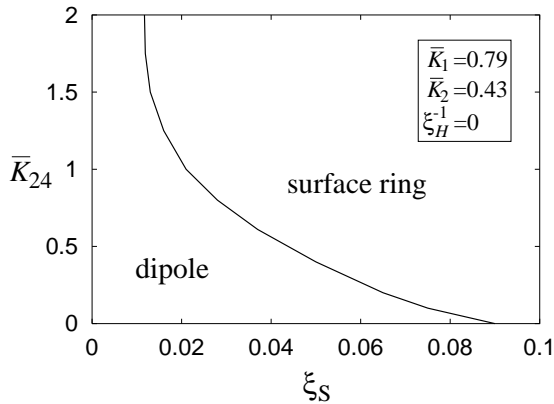,width=7.5cm}}

\vspace{0.2cm}

\caption[]{Phase diagram of the dipole-surface ring transition as a
  function of $\xi_{S}$ and $\overline{K}_{24}$. Further parameters
  are indicated in the inset.}
\label{fig.phase5CBK24}
\end{figure}

For the liquid crystal 5CB we determined the stable configuration as a
function
of $\overline{K}_{24}$ and $\xi_{S}$. The phase diagram is presented
in Fig.\ \ref{fig.phase5CBK24}. With its help we can derive a lower bound for
the surface constant $W$ at the interface of water and 5CB when the
surfactant sodium dodecyl sulfate is involved. As the experiments by 
Poulin\linebreak
\noindent
{\em et al.\/} clearly demonstrate water droplets dispersed in 5CB do 
assume the dipole configuration. From the phase diagram we conclude
$\xi_{S} < 0.09$ as a necessary condition for the existence of the
dipole. With $a\approx 1\,\mu\text{m}$,
$K_{3} = 0.53 \times 10^{-6}\,\mathrm{dyn}$, and the definition\ (\ref{2.10})
for $\xi_{S}$ we arrive at
\begin{equation}
W > 0.6 \, \text{erg}/\text{cm}^{2} \enspace .
\end{equation}
If we assume the validity of the Cauchy-Relation\ (\ref{2.6}) which
for 5CB gives $\overline{K}_{24} = 0.61$, we conclude that 
$W > 1.5 \, \text{erg}/\text{cm}^{2}$. Both values indicate one of the
highest anchoring strengths ever observed.

\section{Conclusions}

The purpose of the article was to give a detailed study of the director
field around a spherical particle and to illustrate 
how it can be manipulated by an
external field. We clearly find that for large particles and
sufficiently strong surface anchoring the dipole is the preferred
configuration. For conventional calamitic liquid crystals where
$K_{2} < K_{1}$ the dipole should always exhibit a twist around
the hyperbolic hedgehog. It should not occur in discotic liquid
crystals where $K_{2} > K_{1}$. According to our calculations the bend
constant $K_{3}$ only plays a minor role for the twist transition.
The Saturn ring appears for sufficiently small particles. However,
the dipole can be transformed into the Saturn ring by means of a
magnetic field if the Saturn ring is metastable at $H=0$. Otherwise a
hysteresis is visible. For the liquid-crystal 5CB we find the Saturn ring
to be metastable for a particle size $a=1\,\mu\text{m}$. Increasing the
radius $a$ this metastability will vanish in analogy with our
calculations within the one-constant approximation (see Fig.\ \ref{fig.size}).
Decreasing the surface-anchoring strength $W$ the surface ring
configuration with a quadrupolar symmetry becomes absolutely stable. We
never find a stable structure with dipolar symmetry where the surface
ring has a general angular position $\theta_{d}$ or is even shrunk to a 
point at $\theta_{d} = 0,\pi$. The surface ring is clearly favored by
a large saddle-splay constant $K_{24}$.

So far convincing experiments on dispersions of spherical particles
in a nematic liquid crystal only exist in the case of inverted nematic
emulsions \cite{itapdb:Poulin1997,itapdb:Poulin1997a,itapdb:Poulin1998}.
We hope that the summary of our results stimulates further experiments
which try different liquid crystals as a host fluid, 
manipulate the anchoring strength, investigate the effect of external
fields, and attempt to disperse silica or latex spheres 
\cite{itapdb:Poulin1994,itapdb:Raghunathan1996,itapdb:Raghunathan1996a}.

\acknowledgements
The author thanks T.~C. Lubensky, P. Poulin, A. R\"udinger, Th. Seitz,
J. Stelzer, E.~M. Terentjev, H.-R. Trebin, and S. \v Zumer for helpful
discussions. The work was supported by the Deutsche
Forschungsgemeinschaft under Grant No. Tr 154/17-1/2.

\appendix
\section{} \label{app.A}
For completeness we give the explicit formulas of 
$\bbox{\nabla} \cdot {\bf n}$, ${\bf n} \cdot \bbox{\nabla} \times {\bf n}$,
and ${\bf n} \times (\bbox{\nabla} \times {\bf n})$ for the
director field in Eq.\ (\ref{2.1}) and the modified spherical
coordinates $(\rho,\theta)$:
\begin{mathletters}
\begin{equation}
\bbox{\nabla} \cdot {\bf n} = \sqrt{\rho}(2 n_{r} - 2 \rho n_{r,\rho}
 + \cot \theta n_{\theta} + n_{\theta,\theta})
\end{equation}
\begin{eqnarray}
{\bf n} \cdot \bbox{\nabla} \times {\bf n} & = &
\sqrt{\rho}(2\rho n_{\theta}n_{\phi,\rho} - 2 \rho n_{\phi} n_{\theta,\rho} 
- n_{\phi} n_{r,\theta} \nonumber \\
 & & + \cot \theta n_{r} n_{\phi} + n_{r}n_{\phi,\theta})
\end{eqnarray}
\begin{eqnarray}
{\bf n} \times (\bbox{\nabla} \times {\bf n}) & = &
\sqrt{\rho} [n_{\theta}(n_{\theta}-2\rho n_{\theta,\rho}-n_{r,\theta})
\nonumber\\
& & + n_{\phi}(n_{\phi}-2\rho n_{\phi,\rho})] {\bf e}_{r} \nonumber\\
& & +\sqrt{\rho} [n_{\phi}(\cot \theta n_{\phi} + n_{\phi,\theta})
\nonumber\\ 
& & - n_{r}(n_{\theta} - 2\rho n_{\theta,\rho} - n_{r,\theta})] 
{\bf e}_{\theta} \nonumber\\
& & -\sqrt{\rho} [n_{r}(n_{\phi} - 2\rho n_{\phi,\rho}) \nonumber\\
& & + n_{\theta}(\cot \theta n_{\phi} + n_{\phi,\theta})] 
{\bf e}_{\phi} \enspace.
\end{eqnarray}
\end{mathletters}
Partial derivatives are indicated by $,i$ ($i=\rho,\theta$) and
\begin{equation}
\frac{\partial}{\partial r} = - 2 \rho^{3/2} \frac{\partial}{\partial \rho}
\end{equation}
was used.

\section{Chain rule for functional derivatives} \label{app.B}
Suppose $F[\Phi_{i}]$ is a functional depending on the fields
$\Phi_{i}({\bf x})$ in real space. The functional derivative 
$\delta F/\delta \Phi_{i}({\bf x})$ is introduced via the
Taylor expansion
\begin{equation}
\label{B.1}
\delta F = F[\Phi_{i} + \delta \Phi_{i}] - F[\Phi_{i}] \approx 
\sum_{i}\int \frac{\delta F}{\delta \Phi_{i}({\bf x})} \, 
\delta \Phi_{i}({\bf x}) \, d^{3}x \enspace.
\end{equation}
$\delta F$ indicates the change of the functional if at every space
point the field $\Phi_{i}({\bf x})$ is changed by the small amount
$\delta \Phi_{i}({\bf x})$. The special choice
$\delta \Phi_{i} = \delta_{ij} \epsilon \delta({\bf x} - {\bf x}_{0})$ leads
directly to the definition
\begin{equation}
\label{B.2}
\left. \frac{\delta F}{\delta \Phi_{j}({\bf x})} \right|_{{\bf x}={\bf x}_{0}}
:= \lim_{\epsilon \rightarrow 0}
\frac{F[\Phi_{i} + \delta_{ij} \epsilon \delta({\bf x} - {\bf x}_{0})]
  -  F[\Phi_{i}]}{\epsilon} \enspace.
\end{equation}
If the field $\Phi_{i}$ depends on other fields $\phi_{k}$, {\em i.e.},
$\Phi_{i}=\Phi_{i}(\phi_{k})$, then with 
\begin{eqnarray}
F[\Phi_{i}(\phi_{k} + \delta_{jk}
\epsilon \delta ({\bf x} - {\bf x}_{0}) )] & \approx & F[\Phi_{i}(\phi_{k})
\nonumber \\
 & &  + \delta_{jk} \epsilon \delta ({\bf x} - {\bf  x}_{0})
   \frac{\partial \Phi_{i}}{\partial \phi_{j}}]
\end{eqnarray}
one obtains immediately the chain rule
\begin{equation}
\frac{\delta F}{\delta \phi_{j}} = \sum_{i} 
\frac{\delta F}{\delta \Phi_{i}} \,
\frac{\partial \Phi_{i}}{\partial \phi_{j}} \enspace
\end{equation}
from Eq.\ (\ref{B.2}).

%\bibliographystyle{prsty}
%\bibliographystyle{aip}
%\bibliography{jourkurz,/user6/holger/paper/emulnum2/paper}

\begin{thebibliography}{10}

\bibitem{itapdb:Russel1995}
W.~B. {Russel}, D.~A. {Saville}, and W.~R. {Schowalter}, {\em Colloidal
  Dispersions} (Cambridge University Press, Cambridge, 1995).

\bibitem{itapdb:Krech1994}
M. {Krech}, {\em The Casimir Effect in Critical Systems} (World Scientific,
  Singapore, 1994).

\bibitem{itapdb:Mostepanenko1997}
V.~M. {Mostepanenko} and N.~N. {Trunov}, {\em The Casimir effect and its
  application} (Clarendon Press, Oxford, 1997).

\bibitem{itapdb:Dinsmore1996}
A.~D. {Dinsmore}, A.~G. {Yodh}, and D.~J. {Pine}, Nature {\bf 383},  259
  (1996).

\bibitem{itapdb:Dinsmore1998}
A.~D. {Dinsmore}, D.~T. {Wong}, P. {Nelson}, and A.~G. {Yodh}, Phys.\ Rev.\
  Lett. {\bf 80},  409  (1998).

\bibitem{itapdb:Rudhardt1998}
D. {Rudhardt}, C. {Bechinger}, and P. {Leiderer}, Phys.\ Rev.\ Lett. {\bf 81},
  1330  (1998).

\bibitem{itapdb:Brochard1970}
F. {Brochard} and P.~G. de~{Gennes}, J.\ Phys.\ (Paris) {\bf 31},  691
  (1970).

\bibitem{itapdb:Ramaswamy1996}
S. {Ramaswamy}, R. {Nityananda}, V.~A. {Raghunathan}, and J. {Prost}, Mol.\
  Cryst.\ Liq.\ Cryst. {\bf 288},  175  (1996).

\bibitem{itapdb:Ruhwandl1997}
R.~W. {Ruhwandl} and E.~M. {Terentjev}, Phys.\ Rev.~E {\bf 55}, 2958 (1997).

\bibitem{itapdb:Poulin1997}
P. {Poulin}, H. {Stark}, T.~C. {Lubensky}, and D.~A. {Weitz}, Science {\bf
  275},  1770  (1997).

\bibitem{itapdb:Lubensky1998}
T.~C. {Lubensky}, D. {Pettey}, N. {Currier}, and H. {Stark}, Phys.\ Rev.~E 
{\bf 57},  610  (1998).

\bibitem{itapdb:Poulin1997a}
P. {Poulin}, V. {Cabuil}, and D.~A. {Weitz}, Phys.\ Rev.\ Lett. 
{\bf  79},  4862 (1997).

\bibitem{itapdb:Poulin1998}
P. {Poulin} and D.~A. {Weitz}, Phys.\ Rev.~E {\bf 57},  626  (1998).

\bibitem{itapdb:Horn1981}
R.~G. {Horn}, J.~N. {Israelachvili}, and E. {Perez}, J.\ Phys.\ (Paris)
  {\bf 42},  39  (1981).

\bibitem{itapdb:Poniewierski1987a}
A. {Poniewierski} and T. {Sluckin}, Liq.\ Cryst. {\bf 2},  281  (1987).

\bibitem{itapdb:Musevic1996}
I. Mu\v sevi\v c, G. Slak, and R. Blinc, Abstract Book p. 91, 16th 
International Liquid Crystal Conference, Kent, USA, 1996.

\bibitem{itapdb:Borstnik1997}
A. {Bor\v stnik} and S. {\v Zumer}, Phys.\ Rev.~E {\bf 56},  3021  (1997).

\bibitem{itapdb:Borstnik1998}
A. Bor\v stnik, H. Stark, and S. \v Zumer, submitted to Phys. Rev. E.

\bibitem{itapdb:Galatola1998}
P. Galatola and J.~B. Fournier, Abstract Book p. O-19, 17th International 
Liquid Crystal Conference, Strasbourg, France, 1998.

\bibitem{itapdb:Ajdari1991}
A. {Ajdari}, L. {Peliti}, and J. {Prost}, Phys.\ Rev.\ Lett. {\bf 66},  1481
  (1991).

\bibitem{itapdb:Swanson1995}
B.~D. {Swanson} and L.~B. {Sorenson}, Phys.\ Rev.\ Lett. {\bf 75},  3293
  (1995).

\bibitem{itapdb:Ziherl1998}
P. {Ziherl}, R. {Podgornik}, and S. {{\v Z}umer}, Chem.\ Phys.\ Lett. 
{\bf 295},  99  (1998).

\bibitem{itapdb:Meyer1972}
R.~B. {Meyer}, Mol.\ Cryst.\ Liq.\ Cryst. {\bf 16},  355  (1972).

\bibitem{itapdb:Kurik1988}
M.~V. {Kurik} and O.~D. {Lavrentovich}, 
Usp. Fiz. Nauk \textbf{154}, 381 (1988) [Sov.\ Phys.\ Usp. {\bf 31},  196
  (1988)].

\bibitem{itapdb:Terentjev1995}
E.~M. {Terentjev}, Phys.\ Rev.~E {\bf 51},  1330  (1995).

\bibitem{itapdb:Kuksenok1996}
O.~V. {Kuksenok}, R.~W. {Ruhwandl}, S.~V. {Shiyanovskii}, and E.~M.
  {Terentjev}, Phys.\ Rev.~E {\bf 54},  5198  (1996).

\bibitem{itapdb:Ruhwandl1997a}
R.~W. {Ruhwandl} and E.~M. {Terentjev}, Phys.\ Rev.~E {\bf 56},  5561  (1997).

\bibitem{itapdb:deGennes1993}
P.~G. de~{Gennes} and J. {Prost}, {\em The Physics of Liquid Crystals}, {\em
  2nd. ed.} (Clarendon Press, Oxford,  1993).

\bibitem{itapdb:Nehring1971}
J. {Nehring} and A. {Saupe}, J.~Chem.\ Phys. {\bf 54},  337  (1971).

\bibitem{itapdb:Blinov1989a}
L.~M. {Blinov}, A.~Y. {Kabayenkov}, and A.~A. {Sonin}, Liq.\ Cryst. {\bf 5},
  645  (1989).

\bibitem{itapdb:Kleman1983}
M. {Kl{\'e}man}, {\em Points, Lines and Walls: In liquid crystals, magnetic
  systems, and various ordered media} (John Wiley \& Sons, New York, 
  1983).

\bibitem{itapdb:Schopohl1987}
N. {Schopohl} and T.~J. {Sluckin}, Phys.\ Rev.\ Lett. {\bf 59},  2582  (1987).

\bibitem{itapdb:Penzenstadler1989}
E. {Penzenstadler} and H.-R. {Trebin}, J.~Phys.\ France {\bf 50},  1027
  (1989).

\bibitem{itapdb:Press1992}
W.~H. {Press}, S.~A. {Teukolsky}, W.~T. {Vetterling}, and B.~P. {Flannery},
  {\em {N}umerical {R}ecipes in {F}ortran: {T}he {A}rt of {S}cientific
  {C}omputing} (Cambridge University Press, Cambridge,  1992).

\bibitem{itapdb:Lavrentovich1986}
O. {Lavrentovich} and E. {Terentjev}, Zh. Eksp. Teor. Fiz. 
{\bf 91}, 2084 (1986) [Sov.\ Phys.\ JETP {\bf 64},  1237 (1986)].

\bibitem{itapdb:Poulin1994}
P. {Poulin}, V.~A. {Raghunathan}, P. {Richetti}, and D. {Roux}, J.~Phys.\ I\
  France {\bf 4},  1557  (1994).

\bibitem{itapdb:Raghunathan1996}
V.~A. {Raghunathan}, P. {Richetti}, and D. {Roux}, Langmuir {\bf 12},  3789
  (1996).

\bibitem{itapdb:Raghunathan1996a}
V.~A. {Raghunathan} {\it et~al.}, Mol.\ Cryst.\ Liq.\ Cryst. {\bf 288},  181
  (1996).

\end{thebibliography}

\end{document}